\documentclass[useAMS,usenatbib]{mn2e}

\usepackage{graphicx}
\def\parsec{\mbox{PARSEC }}
\def\dsed{\mbox{DSED }}

\title[Multichromatic CMDs of NGC\,6366]{Multichromatic 
colour-magnitude diagrams of the globular cluster NGC\,6366\thanks{Based on observations obtained 
at the Southern Astrophysical Research (SOAR) telescope, which is a joint project of the 
Minist\'erio da Ci\^encia, Tecnologia, e Inova\c{c}\~ao (MCTI) da Rep\'ublica Federativa do 
Brasil, the U.S. National Optical Astronomy Observatory (NOAO), the University of North Carolina 
at Chapel Hill (UNC), and Michigan State University (MSU).}\thanks{Based on observations with 
the NASA/ESA Hubble Space Telescope, obtained at the Space Telescope Science Institute, which 
is operated by AURA, Inc., under NASA contract NAS5-26555, under programs GO-10775 (PI: A. 
Sarajedini).}\thanks{Based on data from DENIS Consortium}}

\author[Fab\'iola Campos et al.]{Fab\'iola Campos\thanks{E-mail:
fabiola.campos@ufrgs.br (FC)}, S.O. Kepler, C. Bonatto and J. R. Ducati\\
Departamento de Astronomia, Universidade Federal do Rio Grande do Sul, Av. Bento 
Gon\c{c}alves 9500\\ Porto Alegre 91501-970, RS, Brazil}

\begin{document}

\pagerange{\pageref{firstpage}--\pageref{lastpage}}

\maketitle

\label{firstpage}

\begin{abstract}
We present multichromatic isochrone fits to the colour magnitude data of the globular cluster NGC\,6366, based on HST ACS/WFC and SOAR photometric data. We corrected the photometric data for differential reddening and 
calculated the mean ridge line of the colour magnitude diagrams. We compared the isochrones of Dartmouth Stellar Evolution Database and PAdova and 
TRieste Stellar Evolution Code both with microscopic diffusion starting on the main 
sequence. Bracketing all previous determinations of this cluster 
we tested metallicities from [Fe/H]=--1.00 to [Fe/H]=--0.50, and ages from 9 to 13 Gyrs.
After determining the total to 
selective extinction ratio only from stars belonging to this
cluster, $R_V=3.06\pm0.14$, we found the parameters for this cluster 
to be $E(B-V)= 0.69\pm0.02$(int) $\pm0.04$(ext), 
$(m-M)_V= 15.02\pm0.07$(int) $\pm0.13$(ext), $Age= 11\pm1.15Gyr$.
Evolutionary models fail to reproduce the low--T$_\mathrm{eff}$ sequence in multi--band colour magnitude diagrams, indicating that they still have an incomplete physics. We found that the Dartmouth Stellar Evolution Database isochrones fit better the sub giant branch and low main sequence 
than the PAdova and TRieste Stellar Evolution Code. 
\end{abstract}


\section{Introduction}
\label{intro}

Galactic globular clusters (GCs) are considered to be excellent laboratories for the 
study of stellar evolution, mainly because the stars, in most GCs, follow a
single isochrone, suggesting that they formed roughly at the same time and with the same
metallicity. In the context of Galaxy formation models, accurate astrophysical parameters (e.g. age, metallicity, mass and distance) of GCs are a source of information on Galaxy evolution. In addition, they can also
be used to determine the distance to the center of the Galaxy as done, for example by \citet{bica06}.

Usually, astrophysical parameters of GCs are obtained by fitting models 
to the stars present in colour magnitude diagrams (CMDs). There are at least two free physical parameters, age and metallicity, which are intrinsic to the models, and with two fitting parameters, extinction and distance. In addition, it is necessary to take the uncertainties in the construction of isochrones into account, since they propagate to the derived parameters.

Among the problems associated with the evolutionary models are the lack of a precise description of convection. 
Red giant stars and low mass main sequence have a deep convective envelope, causing a large uncertainty in the models. Besides, stars lose mass (higher rates for massive stars) in the form of stellar wind, and this loss increases several orders of magnitude for stars that already left the main sequence. Theoretically predicting the mass loss rate is very difficult, and the evolutionary models use prescriptions consistent with observations of stars that are at a similar stage. Mass loss depends heavily on metallicity, and this dependence is difficult to measure, creating more uncertainties in the models. It is still necessary to account for the uncertainties associated with opacity,
where the problem is the lack of several molecular species in the opacity tables. This effect is important not only in giant stars, but also at the lower main sequence \citep{bressan13}.

\citet{bolte95} argue that using the colour of the main sequence turn off point (MSTO) must be avoided when 
determining the age of GCs. The MSTO has the highest uncertainty predicted in stellar models, contributing significantly, together with other input parameters, to the total uncertainty. If the MSTO brightness is used to determine the GC age, an uncertainty of 25\% in distance generates an uncertainty of 22\% in age \citep{bolte95}.

On top of the uncertainties in the evolutionary models, there is also the problem of lack of precision in determining the interstellar reddening, as this value varies considerably throughout the Galaxy. The total to selective extinction ratio (R$_V$) depends upon the environment along the line of sight \citep{mathis}, and the distribution of dust and gas in the Galaxy is neither isotropic nor homogeneous, being composed not only of different sizes of dust 
clouds, but also different chemical composition and sizes of dust grains. \citet{mathis} argue that dust properties can vary significantly, even on small angular scales. As an example, \citet{hendricks12} studied the dust properties in the line of sight to the Galactic GC M4. They obtain R$_V$=3.62$\pm$0.07 with the \citet{ccm} extinction law. In fact, they argue that, in former studies, the authors suggest a dust type different than the standard assumption of R$_V$=3.1$\pm$0.05 \citep{wegner}, varying from 3.3 to 4.2, causing 16\% of variation in distance.

\citet{dantona09} studied stars of M\,4 and NGC\,1851, two GCs with similar metallicity and age. They found that the RGB (red giant branch) ``bump" and SGB (sub giant branch) of M\,4 is fainter than in NGC\,1851, which can be explained if the total CNO in M\,4 is higher than in NGC\,1851. This implies that different initial C+N+O abundances between both clusters may lead to differences in the turnoff morphology that can otherwise be attributed to an age difference. 

\citet{marino11,marino12} studied the implications of chemical enrichment and the relative ages of the different populations of M\,22 and $\omega$ Cen respectively. They showed that if they take the observed values of CNO abundances into account in the isochrones, the faint and the bright SGB of M\,22
are almost coeval, and if the effect of C+N+O is not considered, the faint SGB is 1--2 Gyrs older than the bright one. In the case of $\omega$ Cen, they found that the most metal rich population is enhanced by $\approx0.5$ dex in [(C+N+O)/Fe] relative to the most metal poor one. Comparing isochrones with standard and enhanced CNO, they found that the ehnanced ones give younger ages for the same turn off luminosity. They conclude that a trend in CNO/Fe could help reducing the large age spread among the subpopulations of $\omega$ Cen.

\citet{milone12} studied the multiple populations in 47 Tucanae, two of which are clearly present through all evolutionary phases. The authors argue that the most straightforward interpretation of the difference between these two populations is that the one that has less stars is the remnant of the first stellar generation, while the other carries the signatures of CNO and proton--capture processing at high temperatures. The majority population of 47\,Tuc should be considered a second stellar generation, formed from material that was partly processed through stars from the first generation.

The most recent determinations of distance to the Galaxy center from the spatial distribution of the GCs (7.2$\pm$0.3 kpc \citealt{bica06}) and through the orbital period of the star SO--2 around the central black hole (8.4$\pm$0.4 kpc \citealt{ghez08}) are significantly different. This difference can be related to the fact that the value of \citet{bica06} may be underestimated because they used the mean value of R$_V$=3.1 to all GCs, where R$_V$ in different lines of sight in the Galaxy can vary from 2.6 to 7.6 \citep{duca03}. On the other hand the black hole may not be at the center of the mass distribution of the Galaxy (\citealt{merritt} and references therein) and, thus, a precise determination of the distance to the GCs is necessary.

In general, GCs studies are restricted to isochrones fits to a single CMD. Given the uncertainties in the models, it is important to determine the GCs parameters based on multichromatic CMDs, especially with the data at wavelengths wide apart. This would enhance the colour differences among stars and produce the parameters with lower uncertainty. Another important point, as argued by \citet{schiavon12}, is that the last frontier of our growing understanding of the physics of old stellar populations resides in the ultra-violet. In this sense, obtaining evolution models that reproduce multi--colour CMDs simultaneously is extremely necessary. 

This paper is organized as follows: in Sect. \ref{ngc6366} we describe previous works about NGC\,6366. In Sect. \ref{data} we describe the data used in our paper. In Sect. \ref{reduction} we describe the observations, the procedure of reduction and the calibration of the SOAR photometric data. In Sect. \ref{analysis} we analyze and discuss the CMD of the cluster, along with the determination of R$_V$ in the line of sight of NGC\,6366. Results and concluding remarks are given in Sect. \ref{results}.

\section{NGC\,6366}
\label{ngc6366}
NGC\,6366 is a relatively open globular cluster, currently ranked as the fifth nearest the Sun. It is located near the disc and is fairly rich in metals. 

\citet{zinn85} classified the metal-poor GCs with slow rotational velocity, high velocity dispersion (V$_{rot}\approx$50$\pm$23 km/s and V$_\sigma$=114$\pm$9 km/s) and distribution essentially spherical around the Galactic center as belonging to the halo. The metal-rich GCs with a more flatten distribution around the Galactic center, with fast speed and lower velocity dispersion (V$_{rot}\sim$152$\pm$29 km/s and V$_\sigma$=71km/s) have been rated as belonging to the bulge and disk of the Galaxy \citep{zinn85}. Later \citet{barbuy98} classified the metal-rich GCs as belonging to the bulge.

For NGC\,6366 \citet{ds89} argue that its kinematic parameter (V$_\sigma$=125$\pm$13 km/s) is incompatible with the disc, which led it to be classified as belonging to the halo system.

According to \citet{harris93}, \citet{rose00} and \citet{sarajedini}, NGC\,6366 has received little attention due to its low central concentration, its projection near the Galactic core direction of the Galaxy ($\ell=18.41\degr$) and low Galactic latitude ($b=16.40\degr$), its high extinction and differential reddening.

The most recent V$\times$B-V CMD in the literature was published by \citet{alonso97}. The data were obtained with the 2.5 Isaac Newton telescope and covered a field of 12.1$\arcmin\times 12.1\arcmin$, reaching the magnitude V$\sim$20.5. They found E(B-V)=0.70$\pm$0.05 and (m-M)$_0$=12.26$\pm$0.15. They show the presence of differential reddening in NGC\,6366, because the northern half of the stars are fainter and redder than the southern half, resulting in a difference of $\Delta$E(B-V)=0.03 between the stars from both sides.

\citet{sarajedini} published data obtained with the ACS/HST. They reached about seven magnitudes below the MSTO, i.e. $m_{F606W}\sim$25.8. Fitting Dartmouth Stellar Evolution Database models [DSED - \citep{dottergc}] with [Fe/H]=--0.73 and [$\alpha$/Fe]={+0.2} and ages between 10 and 14 Gyr, they found a reddening of E(F606W-F814W)=0.70 and (m-M)$_0$=12.69.

\citet{paust09} used the same data published by \citet{sarajedini}, finding a reddening of E(F606W-F814W)=0.76 and $(m-M)_0$=12.6, fitting DSED models \citep{dotter07} with 13.5 Gyr and [Fe/H]=--0.85. However, the main goals in the analysis from \citet{paust09} were the luminosity and mass functions. 

\citet{marin09}, by measuring relative ages, classified NGC\,6366 as an old globular cluster with relative ages ranging from 12.16 Gyr to 13.30 Gyr, depending on metallicity scales. More recently \citet{dotter10}, performing isochrone fitting, also with HST/ACS data obtained by \citet{sarajedini}, estimated the age of NGC\,6366 as 12.00$\pm$0.75 Gyr with [Fe/H]=--0.70 and [$\alpha$/Fe]=+0.2.

\begin{table*}
\centering
\caption[]{Some previous determinations of NGC\,6366 parameters}
\label{tab2}
\begin{minipage}{126mm}
\begin{tabular}{@{}lcccc}
\hline
\textbf{Author} & $ \textbf{[Fe/H]} $ & $ \textbf{E(B-V)} $ & $ \textbf{Distance (kpc)} $ & $ \textbf{Age (Gyr)}$\\
\hline
\citealt{harris93} & --0.78 & 0.80 & 3.00 & --- \\
\citealt{harris96} (2010 edition) & --0.59 & 0.71 & 3.50 & --- \\
\citealt{alonso97} & --0.80 & 0.70$\pm$0.05 & 2.80 & --- \\
\citealt{sarajedini} & --0.73 & 0.75 & 3.75 & 12--14 \\ 
\citealt{paust09} & --0.85 & 0.78 & 3.31 & 13.5 \\ 
\citealt{dotter10} & --0.70 & 0.73 & 3.52 & 12.00$\pm$0.75 \\ 
\hline
\end{tabular}

\medskip 
{\em Distance} calculated considering R$_V$=3.1 \citep{wegner}.
\end{minipage}
\end{table*}

\section{Photometric Data}
\label{data}

The optical ground data on NGC\,6366 discussed in this work were obtained with the SOAR telescope in 2005, 2009 and 2010. The images were centered at the geometric center and have 2048$\times$2048 pixels, with a resolution of
0.153$\arcsec$/pixel, covering 5$\arcmin$x5$\arcmin$. The exposure times for each filter were: 6$\times$(1800s) for U; 5$\times$(30s), 2$\times$(300s) and 1$\times$(1800s) for B, and 5$\times$(30s), 2$\times$(300s) and 2$\times$(1800s) for V. To minimize saturation for the brightest stars, images with short exposure times were obtained only for B and V bands.

The HST ACS/WFC photometric data were obtained from  http://www.astro.ufl.edu/$\sim$ata/public\_hstgc/.
The data are part of the HST treasury program \textquotedblleft{An ACS Survey of 
the Galactic Globular Clusters}\textquotedblright (GO10775 P.I. Ata Sarajedini, \citealt{sarajedini}).
The images are centered at the cluster core, covering 3$\arcmin$x3$\arcmin$. Each photometric band (F606W and F814W) was observed in one orbit, with one short exposure image (10.4s) and four long exposures (140s) for each filter.

To determine the interstellar R$_V$ of cluster stars, we use UBVRI data from Stetson standard stars (\citealt{stetson00}; \citealt{stetson05}) and infrared (J and K) data from the third DENIS release \citep{borse06}.

\section{Data Reduction}
\label{reduction}

After the basic pre-processing steps, over-scan correction and trimming, bias removal, dark current removal, and flat-fielding, we performed the photometry on the SOAR data, with the software DAOPHOT \citep{stetson}. The main parameters necessary to find the stars in the image are the standard deviation ($\sigma$, in counts), the full width at half maximum (FWHM) and the detection limit (threshold, in multiples of $\sigma$). A very low threshold can cause the detection of non-star sources, just random fluctuations of background but, a very high value would fail to detect faint sources. The values of the FWHM for each of our images are listed in Table \ref{tab1}.

Although NGC\,6366 is relatively open and our images cover only approximately one third of its size, the images still present some source confusion, mainly because, in the long exposure images, the bright stars are saturated and can fuse with nearby stars. To minimize this effect, we performed the photometry by fitting the point spread function (PSF).

To calculate the PSF we choose, on average, 35 stars per image. These stars are below saturation, do not have close neighbors, high noise or detector defects near them. We tested all the functions provided by DAOPHOT to calculate the PSF [Gaussian, Lorentzian, Moffat ($\beta$=1.5, 2.5 and 3.5), Penny1 and Penny2]. The software calculates the parameters of the functions and the $\chi$ (defined as the mean square root of the residuals). The function that produced the lowest $\chi$ was Moffat with $\beta$=3.5, that with the steepest peak.

One of the output parameters of the PSF photometry is \textit{sharpness}, related to the intrinsic angular size of the object. For an isolated star, \textit{sharpness} must have a value close to zero; but for semi-resolved galaxies and double sources, \textit{sharpness} will be significantly higher than zero. For cosmic rays and image defects, the value of \textit{sharpness} will be much lower than zero. We considered that only objects with $\left|\textit{sharpness}\right|\leq$2.0 are stars. We detected 2609 stars in the V$\times$B-V CMD, reaching 3.5 magnitudes bellow the MSTO, and 2284 stars in the V$\times$U-V, reaching 2.5 magnitudes below the MSTO. The latter CMD has less stars than the former because the U band has higher extinction than the other photometric bands.

After obtaining the instrumental magnitudes, we performed a calibration to the standard photometric system. We obtained images of the field around the photometric standard Mark A on the same night of the data observations in two different air masses for bands B and V. We performed the photometry on the images of the standard field using the same methods applied to NGC\,6366. So, with the magnitudes for the standard stars, it was possible, through least squares, to determine the equations that transform the instrumental magnitudes to the standard system. We used standard stars from the cluster field itself from the catalog of Peter Stetson, found at $http://www3.cadc-ccda.hia-iha.nrccnrc.gc.ca/community/STETSON/standards/$. We could not use only stars from the cluster itself to do the calibration, because most of the Stetson standard stars in common with the field of NGC\,6366 observed by SOAR were in the RGB, and this would cause uncertainty in the calibration of bluer stars.

We also corrected the photometric data (SOAR and HST ACS/WFC) for differential reddening. We started by dividing the WFC/ACS field of view across NGC\,6366 in a regular grid of $13\times13$ ($13.8\arcsec\times13.8\arcsec$) cells along right ascension and declination, so that the minimum number of stars in each cell is 50.  Next, we selected a subsample of stars containing probable members (with colours compatible with the cluster sequence) having low to moderate colour uncertainty. Then, the individual Hess diagrams built from CMDs extracted in all
cells are matched to the mean (containing all the probable member stars available in the image) one, by shifting the apparent distance modulus and colour excess along the reddening vector by amounts related to the reddening value e(B-V) according to the absorption relations in \citet{ccm}. Since a differentially-reddened cluster should contain cells bluer and redder than the mean, this procedure is equivalent to computing the reddening dispersion around the mean. The last step is to calculate the difference in e(B-V) between all cells and the bluest one, thus yielding the cell to cell distribution of $\delta$E(B-V), from which we compute the mean and maximum values 
occurring in the GC, $\langle\delta$E(B-V)$\rangle$ and $\delta$E(B-V)$_{max}$, respectively \citep{bonatto13}.

\begin{figure}
\resizebox{\hsize}{!}{\includegraphics[clip=true]{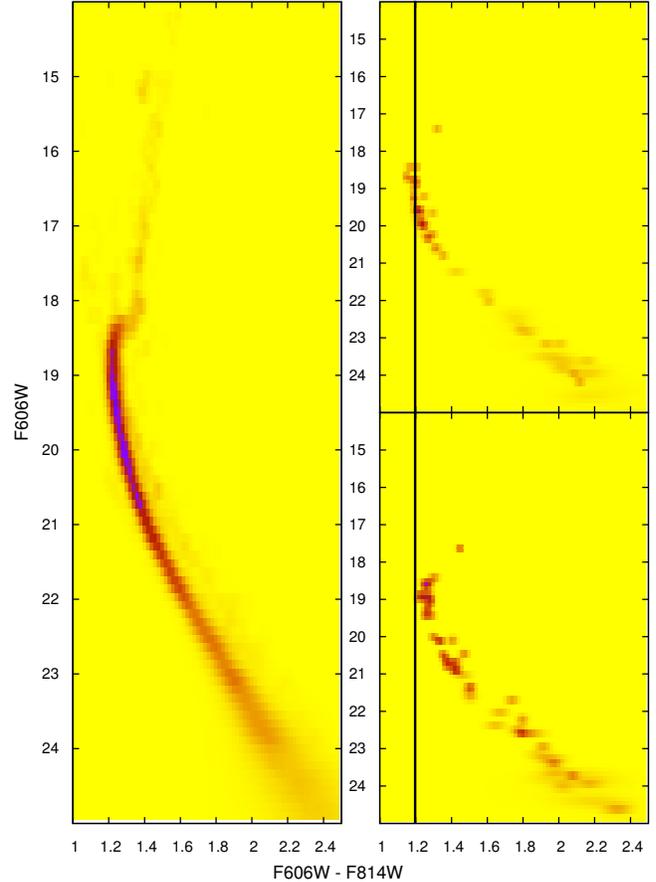}}
\caption{\protect
\footnotesize
The bluest (top-right panel) and reddest (bottom-right) Hess diagrams of NGC\,6366. The difference in reddening between both amounts to $\delta$E(B−V)=0.112 (Shown as the vertical line). The left panel shows the observed average Hess diagram.
}
\label{redd}
\end{figure}

\begin{figure}
\resizebox{\hsize}{!}{\includegraphics[clip=true]{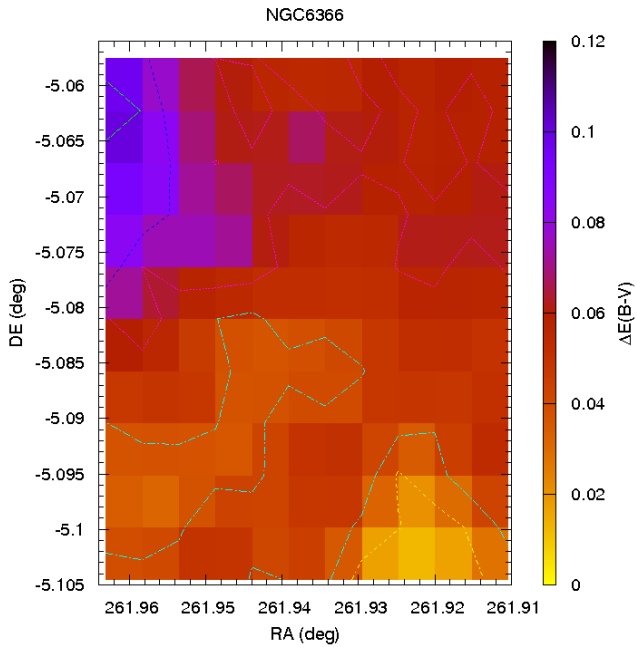}}
\caption{\protect
\footnotesize
The reddening map for NGC\,6366. The stars at the north appear fainter and redder than stars in the south.
}
\label{map}
\end{figure}

We found that the maximum difference between the Hess diagram of the cells are $\delta$E(B-V)=0.112 (figure \ref{redd}) and the mean differential reddening for NGC\,6366 is $\langle\delta$E(B-V)$\rangle$=0.055$\pm$0.018, similar to the $\delta$E(B-V)$\sim$0.03 estimated by \citet{alonso97}. With this analysis we could generate the reddening map for NGC\,6366 that is shown in figure \ref{map}. It is possible to notice that the stars at the north appear fainter and redder than stars in the south, in agreement with \citet{alonso97}.

\begin{table}
\centering
\caption[]{FWHM of all our SOAR images}
\label{tab1}
\renewcommand{\tabcolsep}{2.1mm}
\begin{tabular}{lcc}
\hline
\textbf{Band} & $ \textbf{Exposure Time} $ & $\textbf{FWHM}$\\
(1) & (2) & (3)\\
\hline
U & 6$\times$(1800) & 1.05 \\
B & 1800 & 1.12 \\ 
B & 2$\times$(300) & 0.92 \\ 
B & 5$\times$(30) & 1.07 \\ 
V & 2$\times$(1800) & 1.37 \\ 
V & 2$\times$(300) & 0.95 \\ 
V & 5$\times$(30) & 1.01 \\ 
\hline
\end{tabular}
\begin{list}{Table Notes.}
\item Col.~1: Photometric Band; Cols.~2: Exposure time in seconds; Col.~3: Full width at half maximum ($\arcsec$)
\end{list}
\end{table}

\section{Data Analysis}
\label{analysis}

\subsection{Isochrone Fitting}
\label{fitting}
\citet{milone12} measured the fraction of binaries and the distribution of the GCs, including NGC\,6366, observed by the HST WFC/ACS as a part of the Globular Cluster Treasury project. They estimated that the fraction of binary stars for NGC\,6366 is $f_{bin}^{TOT}=$0.184$\pm$0.014. To take the effects of binarity and also the photometric scatter into account, we calculated the mean ridge line (MRL) of each CMD. The MRL was determined by calculating the mean value of colour and magnitude of bins of 0.04 mag excluding the binary stars.

\citet{dotter08} compared their isochrone models with those from BaSTI \citep{pie04, pie06, cordier}, Padova \citep{girardi00}, Victoria-Regina \citep{vande06} and Yale-Yonsei \citep{yi03, yi04}. They showed that there is a general agreement among the different sets, except for Padova isochrones that appeared hotter and bluer on the lower main sequence and cooler and redder near the MSTO and on the RGB. Yale-Yonsei, Victoria-Regina and BaSTI show differences near the MSTO, where the adopted core overshooting treatments differ and on the lower main sequence where the adopted equation of state and minimum masses differ. 

\citet{jofre} explored the effect of atomic diffusion in the resulting ages of halo metal--poor stellar populations and found an absolute difference of 4 Gyr for ages obtained ignoring or including atomic diffusion in the stellar models. They also tested the age using BaSTI \citep{cass,pie04} and the Yonsei--Yale isochrones \citep{yi03, yi04}, as examples of isochrones without and with diffusion, respectively. They found that ages obtained with BaSTI models agreed with those obtained with isochrones without atomic diffusion, while Yonsei–Yale results agreed better with isochrones with atomic diffusion. \citet{jofre} argue that metal--poor halo stars would be older than the Universe if the atomic diffusion is fully inhibited in the models.

Taking the analysis of \citet{jofre} into account we performed the isochrone fitting to the MRL of the three CMDs (V$\times$B-V, V$\times$U-V and F606W$\times$F606W-F814W) of NGC\,6366 with DSED (\citealt{dotter08}, version 2012) and PAdova and TRieste Stellar Evolution Code (\parsec - \citealt{bressan12}), both including atomic diffusion starting on the main sequence.

We considered the full range of metallicity previously determined for this cluster, which is --1.0$<$[Fe/H]$<$--0.5. \citet{pike}, by computing the S parameter \citep{hartwick} to his photometric data and adopting E(B-V)=0.72$\pm$0.02, estimated [M/H]=--0.50$\pm$0.2. From TiO band photometry of seven giants, \citet{johnson} derived [Fe/H]=--0.6$\pm$0.2. \citet{zw84} recalibrated the TiO band–strength indices onto their own abundance system and derived [Fe/H]=--0.99$\pm$0.25. \citet{ds89}, from spectra obtained of the CaII infrared triplet of several giants belonging to the cluster, estimated [Fe/H]=--0.85; more recently \citet{dacosta95} reanalyzed the same data and estimated [Fe/H]=--0.67.

Among the age estimates of NGC\,6366 found in the literature is the analysis by \citet{alonso97}, comparing
the MRL of this cluster to the ones of NGC 6171 (old and metal rich GC) and Pal12 (youngest known GC), both with [Fe/H]$\approx$--0.80. They found that the age of NGC\,6366 is very similar to that of NGC 6171, and Pal 12 is 4-5Gyr younger. \citet{rose99} estimated an age of about 11 Gyr, using a data base of 34 GCs, through the analysis of relative ages by the magnitude difference $\Delta$V$_{TO}^{HB}$ between the horizontal branch and the MSTO, and the color difference $\delta$(V--I)$_{@2.5}$ between the MSTO and the red giant branch (where the RGB color is measured 2.5 magnitudes above the TO).
\citet{salaris02}, using a group of GCs whose ages they estimate to be well determined (M15, M3, NGC 6171
and 47 Tuc), give an age of 9.5$\pm$1.4 Gyr for NGC\,6366, with the same method applied by \citet{rose99}. To bracket all previous determinations, we used models with ages ranging from 9 to 13 Gyr.

\begin{figure*}
\resizebox{\hsize}{!}{\includegraphics[clip=true]{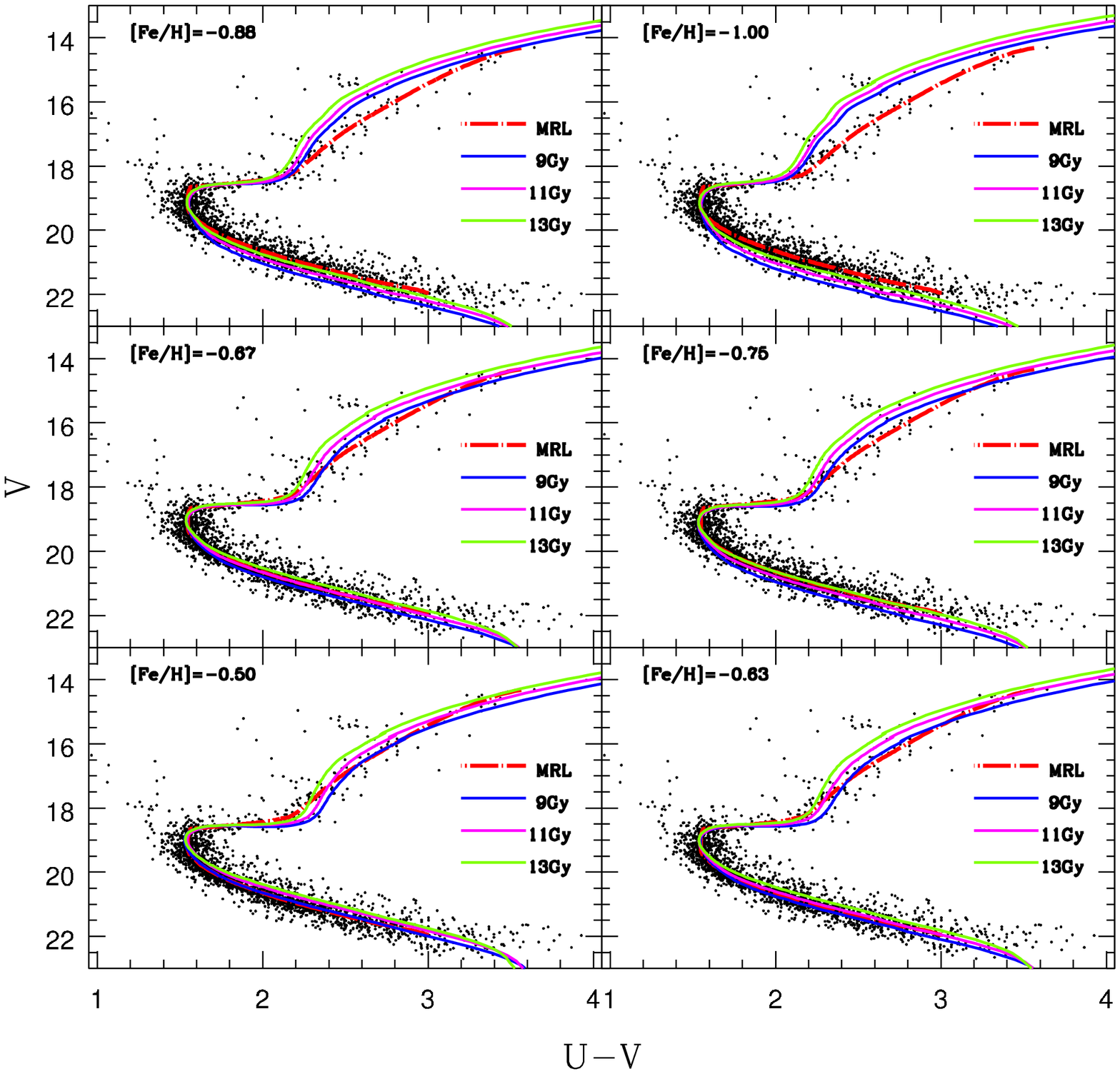}}
\caption{\protect\footnotesize
\dsed fits to the MRL (red dash-dotted line) of NGC\,6366, at U-V colour, considering 
full range of metallicity determinations for this cluster and ages of 9 Gyr (blue), 11 Gyr (magenta) and 13 Gyr (green). The best fit occurs for [Fe/H]=--0.67, consistent with the most recent spectroscopic determination by \citet{dacosta95}.
}
\label{all}
\end{figure*}

In Fig. \ref{all} we show our CMD of the V$\times$U-V, the first ultraviolet CMD for this cluster, together with the fit of DSED models to the MRL, performed by visual inspection. It is not difficult to notice, by looking at the low main sequence and the subgiant branch that, as the 
metallicity decreases (bottom to the top), the models fit the data better, until [Fe/H]=--0.67, when the best fit is found. This model is the best fit in all sectors of the CMD, having a metallicity consistent with the spectroscopic determination by \citet{dacosta95} of RGB stars in the cluster. As metallicity continues to decrease, the models no longer fit the data. We also performed the analysis to the three colours with PARSEC models, finding similar results.

\begin{figure*}
\resizebox{\hsize}{!}{\includegraphics[clip=true]{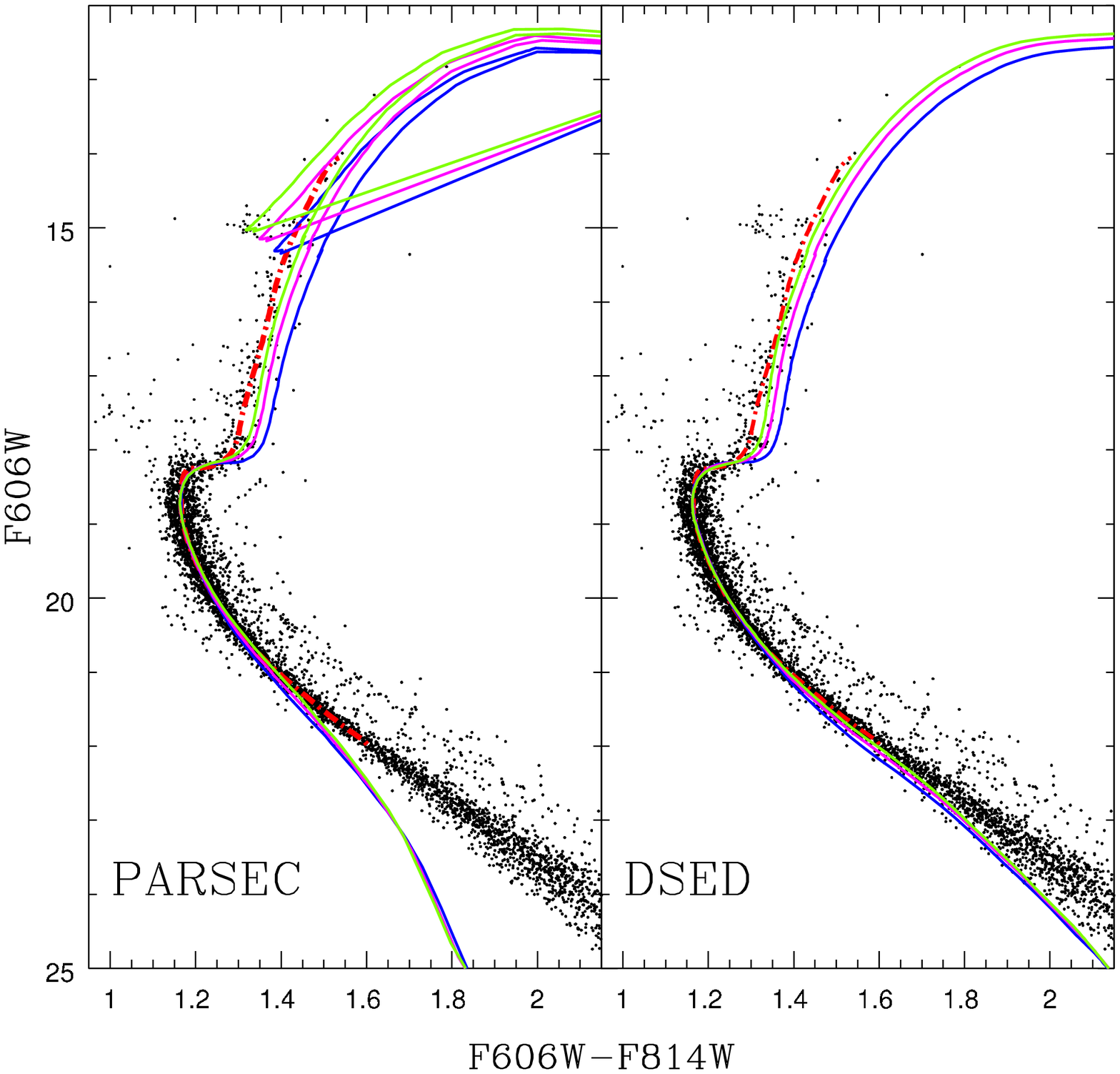}}
\caption{\protect\footnotesize
\parsec and \dsed fits to the MRL (red dash-dotted line) 
of the ACS/HST data, with [Fe/H]=--0.67$\pm$0.07 \citep{dacosta95} 
and ages of 9 Gyr (blue), 11 Gyr (magenta) and 13 Gyr (green). 
\dsed fits the subgiant branch and the lower main sequence better than \parsec.
}
\label{acsfits}
\end{figure*}

An important point in Fig. \ref{acsfits} is that DSED models fit better the lower main sequence and the SGB than PARSEC; this is possibly related to the equation of state adopted by \citet{dotter08}
for stars with mass lower than 0.80M$_\odot$. For that reason, we use DSED to determine the parameters of the cluster.

\begin{figure*}
\resizebox{\hsize}{!}{\includegraphics[clip=true]{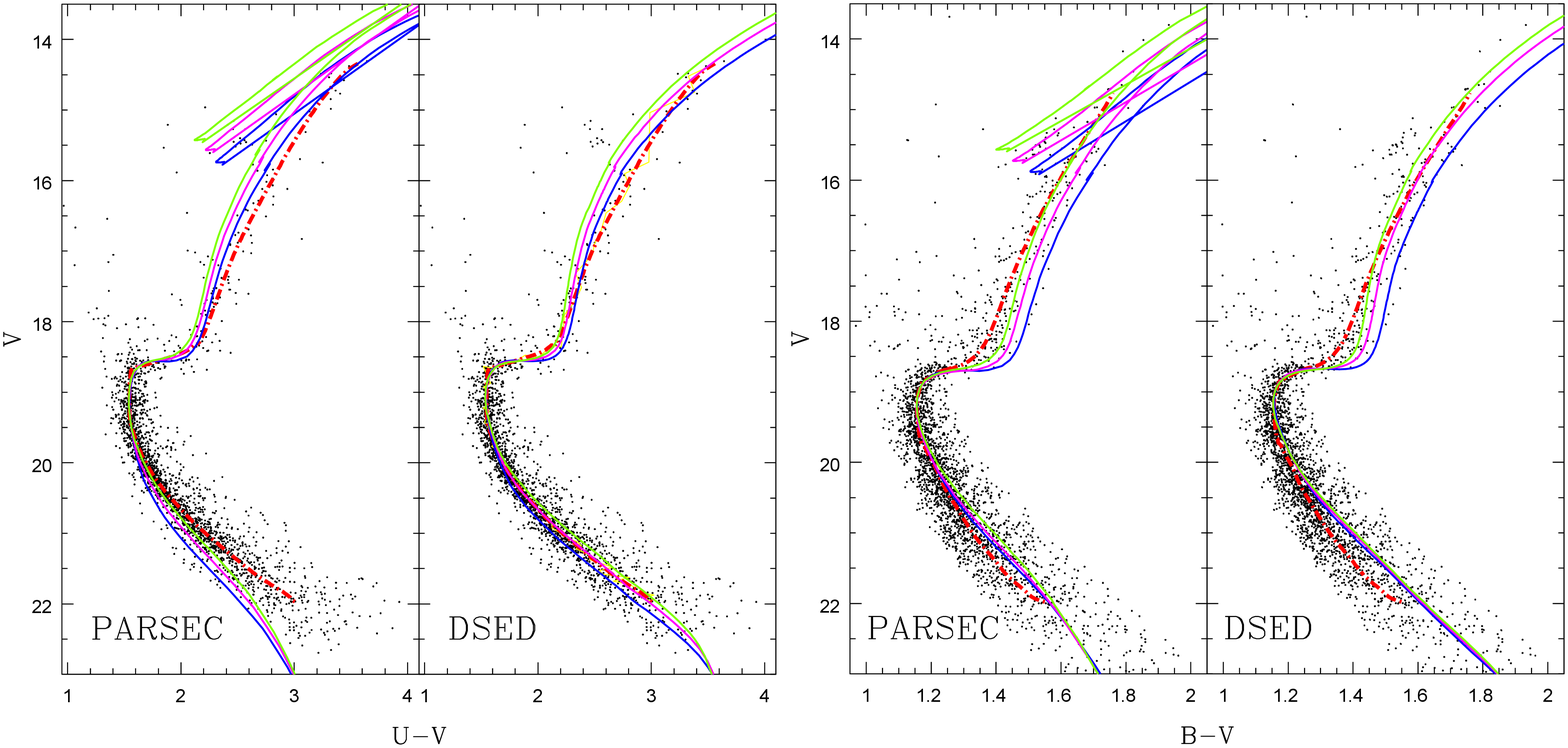}}
\caption{\protect\footnotesize
Same as Fig. \ref{acsfits} for SOAR data. The uncertainties do not decrease when we fit simultaneous colours, because a single model does not fit the three colours simultaneously.} 
\label{fits}
\end{figure*}

The DSED fit for the three CMDs with [Fe/H]=--0.67$\pm$0.07 (Figs. \ref{acsfits} and \ref{fits}) shows that, while the best fit to U-V occurs with 9 Gyr, for B-V it is around 11 Gyr, and for F606W-F814W, with 13 Gyr. The PARSEC best fit to U-V and F606W-F814W occurs at the same ages as DSED. For B-V, the best PARSEC fit occurs also around 13 Gyr. With that, we find the age as 11.00$\pm$1.15 Gyr, with DSED models.

Even when we fit more than one colour, a single model does not fit the three colours simultaneously. The evolutionary models have improved over the years, but they still do not fit the data well, mostly when we consider the red giant branch. This indicates that the evolutionary models still have unsolved problems, such as convection, opacity tables, bolometric corrections and colour effective temperature (T$_\mathrm{eff}$) relations to transform theoretical quantities (luminosity and T$_\mathrm{eff}$) to magnitudes and colours.

\citet{pie09} argues that if the total CNO abundance is the same as in the reference $\alpha$--enhanced composition, one can safely use $\alpha$--enhanced isochrones to represent GC subpopulations affected by CNONa anticorrelations. To take \citet{pie09} arguments into account, we performed the fitting of DSED models with different values of $\alpha$--enhancement, but no improvement in the isochrone fittings were found.

\subsection{Total to Selective Extinction Ratio}
\label{extinction}
For a precise distance, we must take the interstellar reddening into account. So, we determined R$_V$ for NGC\,6366, by applying the method developed by \citet{duca03} with stars belonging to the cluster itself. With this method, it is possible to determine simultaneously the total absorption A$_V$ and the relative R$_V$ from the fit of the extinction law by \citet{ccm} to the photometry of the stars.

To perform this analysis we used NGC\,6366 stars with UBVRI data from Stetson standard stars and JK data from the DENIS third release. Then, we classified the stars as members of NGC\,6366 by building the CMD in different colours. Each star that consistently occurs in the cluster evolutionary sequence in all CMDs, within the photometric uncertainties, is considered a cluster member. To classify the spectral type of the stars, we used \citet{alonso99} tables of the effective temperature of giant stars (F0-K5) for different colours, taking metallicity into account, considering all the range of uncertainties of NGC\,6366 (0.65$<$E(B-V)$<$0.73 and --1.0$<$[Fe/H]$<$--0.50). Since \citet{alonso99} tables are only for giant stars, we used only RGB stars in our analysis, and our sample ended up with 29 stars.

We applied the method developed by \citet{duca03}, fitting the extinction curves to the colours of member stars, obtaining R$_V$=3.06$\pm$0.14 for NGC\,6366. This value is comparable to the mean value for the Galaxy of \citet{wegner}, but it was determined with stars from the cluster itself, and represents the integral extinction through the line of sight all the way to the cluster.

\section{Results and Conclusions}
\label{results}
We show that DSED models lead to a better fit to the data than PARSEC, specially in the SGB and the low main sequence, possibly related to the equation of state adopted by \citet{dotter08} for stars with low mass and the opacity tables for cooler stars.

The mean value of the differential reddening that we found in an area covering 3$\arcmin$x3$\arcmin$ ($\langle\delta$E(B-V)$\rangle$=0.055$\pm$0.018) is close to the estimate by \citet{alonso97} covering 12.2$\arcmin$x12.2$\arcmin$ ($\delta$E(B-V)$\sim$0.03), but our method takes the difference between each cell in the cluster into account, and shifts these stars to the mean diagram along the reddening vector; while \citet{alonso97} calculated the fiducial lines for four halves of the cluster (i.e northern, southern, eastern and western halves) and, then compared de resulting lines on the CMD, finding that there is no difference between the mean lines of the west and east, but stars at the north are fainter and redder than stars in the south.

We also determined the total to selective reddening ratio for NGC\,6366, finding $R_V=3.06\pm0.14$, determined with probable member stars. With that, we estimated the relations E(B-V)=1.02E(F606W-F814W) and E(B-V)=0.57E(U-V) using \citet{ccm}, and determined foreground reddening, distance and age of NGC\,6366 as: 

E(B-V)= $0.69\pm0.02(int)\pm0.04(ext)$;  

d=[$3.82\pm0.15(int)\pm0.01(ext)$]kpc;  

Age= 11.00$\pm$1.15 Gyr; \\
where \textquotedblleft int\textquotedblright means internal uncertainties, i.e. the uncertainties related to the measurements and calibrations; and \textquotedblleft ext\textquotedblright the external ones, i.e. those related to the model fittings. 

Our values of age, distance and foreground reddening for NGC\,6366 seem to agree with the recent determinations of \citet{sarajedini} and \citet{paust09}, but our quoted values ​​of the uncertainties in the parameters are larger. However, unlike previous authors, we consider all terms of uncertainty in all the steps, including the R$_V$ determination with member stars, not the mean value of the Galaxy, as commonly used to determine the parameters of GCs. Another important point is that we take the effect of uncertainties due to the the models into account, through the fit to multiple colours.

The age determined for NGC\,6366 in our analysis could be overestimated if the cluster is CNO enhanced. However, so far there is no C+N+O determination for this cluster. Our results could be affected by the presence of multiple populations, detected by \citet{monelli13} with a new photometric index c$_{U,B,I}$= (U--B)--(B--I) in NGC\,6366's RGB.

One important conclusion is that isochrone fit uncertainties to NGC\,6366 do not decrease when we use multiple colours, because a single model does not fit the three colours simultaneously. Clearly, models fail to reproduce the observed evolutionary sequences of NGC\,6366 (especially for the low--T$_\mathrm{eff}$ stars) in CMDs built with photometry obtained with different instrumentation such as SOAR and ACS/HST, even if we fit isochrone models with $\alpha$--enhancement. This indicates that isochrone models still have problems that remain unsolved, possibly convection, opacity tables, bolometric corrections and colour--effective temperature relations.

With this work, we determined realistic uncertainties in the parameters of NGC\,6366. This kind of analysis must be extended to other GCs for a more precise determination of fundamental parameters of the Galaxy.

\section*{Acknowledgments}
We thank an anonymous referee for important comments and suggestions. Partial financial support for this research comes from CNPq and PRONEX--FAPERGS/CNPq (Brazil).
HST data used in this paper were directly obtained from {\em The ACS Globular Cluster Survey: 
 http://www.astro.ufl.edu/$\sim$ata/public\_hstgc/}.


\end{document}